\documentclass[aip,reprint,pof]{revtex4-1}

\usepackage{graphicx} %eps figures can be used instead
\usepackage{amsmath}

\newcommand{\mupiv}{$\mu$-PIV}
\begin{document}

% Use the \preprint command to place your local institutional report number 
% on the title page in preprint mode.
% Multiple \preprint commands are allowed.
%\preprint{}
\title{
Experimental evidence of slippage breakdown for a superhydrophobic
surface in a microfluidic device
} %Title of paper

% repeat the \author .. \affiliation  etc. as needed
% \email, \thanks, \homepage, \altaffiliation all apply to the current author.
% Explanatory text should go in the []'s, 
% actual e-mail address or url should go in the {}'s for \email and \homepage.
% Please use the appropriate macro for the type of information

% \affiliation command applies to all authors since the last \affiliation command. 
% The \affiliation command should follow the other information.
%G. Bolognesi,\textit{$^{a}$}
%C. Cottin-Bizonne\textit{$^{a}$} and $^{\ast}$\textit{$^{a,\dagger}$} C. Pirat

\author{G. Bolognesi}
\email[]{g.bolognesi@imperial.ac.uk}
%\homepage[]{Your web page}
%\thanks{}
\altaffiliation{Present address: Dept. of Chemistry, Imperial College London, South Kensington Campus, SW7 2AZ, London UK }
\affiliation{ILM, Universit\'e de Lyon, Universit\'e de Lyon 1 and CNRS, UMR5306, F-69622 Villeurbanne, France}

% Collaboration name, if desired (requires use of superscriptaddress option in \documentclass). 
% \noaffiliation is required (may also be used with the \author command).
%\collaboration{}
%\noaffiliation
\author{C. Cottin-Bizonne}
%\email[]{}
%\homepage[]{Your web page}
%\thanks{}
\affiliation{ILM, Universit\'e de Lyon, Universit\'e de Lyon 1 and CNRS, UMR5306, F-69622 Villeurbanne, France}
\author{C. Pirat}
%\email[]{}
%\homepage[]{Your web page}
%\thanks{}
\affiliation{ILM, Universit\'e de Lyon, Universit\'e de Lyon 1 and CNRS, UMR5306, F-69622 Villeurbanne, France}
%\author{L. Bocquet}
%\email[]{}
%\homepage[]{Your web page}
%\thanks{}
%\affiliation{ILM, Université de Lyon, Université de Lyon 1 and CNRS, UMR 5306, F-69622 Villeurbanne, France}

\date{\today}

\begin{abstract}
% insert abstract here
A full characterization of the water flow past a silicon superhydrophobic
surface with longitudinal micro-grooves enclosed in a microfluidic device 
is presented. Fluorescence microscopy
images of the flow seeded with fluorescent passive tracers were digitally
processed to measure both the velocity field and the position
and shape of the liquid-air interfaces at the superhydrophobic surface.  The
simultaneous access to the meniscus and velocity profiles allows us to put
under a strict test the no-shear boundary condition at the liquid-air
interface.  Surprisingly, our measurements show that air pockets in
the surface cavities can sustain non-zero interfacial shear stresses, thereby
hampering the friction reduction capabilities of the surface.
The effects of the meniscus position and shape as well as 
of the liquid-air interfacial friction 
on the surface performances are separately
assessed and quantified.

\end{abstract}

% insert suggested PACS numbers in braces on next line

\maketitle %\maketitle must follow title, authors, abstract and \pacs

% Body of paper goes here. Use proper sectioning commands. 
% References should be done using the \cite, \ref, and \label commands
%\section{}
%\label{}
%\subsection{}
%\subsubsection{}

% If in two-column mode, this environment will change to single-column format so that long equations can be displayed. 
% Use only when necessary.
%\begin{widetext}
%$$\mbox{put long equation here}$$
%\end{widetext}

% Figures should be put into the text as floats. 
% Use the graphics or graphicx packages (distributed with LaTeX2e).
% See the LaTeX Graphics Companion by Michel Goosens, Sebastian Rahtz, and Frank Mittelbach for examples. 
%
% Here is an example of the general form of a figure:
% Fill in the caption in the braces of the \caption{} command. 
% Put the label that you will use with \ref{} command in the braces of the \label{} command.
%
% \begin{figure}
% \includegraphics{}%
% \caption{\label{}}%
% \end{figure}

% Tables may be be put in the text as floats.
% Here is an example of the general form of a table:
% Fill in the caption in the braces of the \caption{} command. Put the label
% that you will use with \ref{} command in the braces of the \label{} command.
% Insert the column specifiers (l, r, c, d, etc.) in the empty braces of the
% \begin{tabular}{} command.
%
% \begin{table}
% \caption{\label{} }
% \begin{tabular}{}
% \end{tabular}
% \end{table}

\section{Introduction}

In lab-on-a-chip technology, the high hydrodynamic resistance the liquid
experiences as it flows in a network of micrometric channels often imposes the
microfluidic device to be connected to pumping systems thousands of times
bigger than the chip itself, thereby erasing many advantages of a
millimeter-sized device.  In this respect, superhydrophobic surfaces (SHS) have
demonstrated to be an effective tool to relax the constraint of no-moving
liquid at the channel walls\cite{rothstein2010slip}, due to air pockets trapped
in surface cavities (the so-called Cassie or Fakir state). Air bubbles durably reside
in the surface roughness only under specific conditions of pressure and
surface wettability\cite{cottin2003low}. When the Cassie state is no longer stable, 
the meniscus collapses within the cavities\cite{peters2009cassie,giacomello2012cassie,giacomello2012metastable}
and the liquid is in contact with the entire exposed surface
of the solid (the so-called Wenzel state). 
%In the last
%decades, researchers have been inspired by the plant and animal worlds to
%device and fabricate artificial
%SHS\cite{mock2005towards,otten2004plants,gao2004biophysics,parker2001water}.
%At the same time, several
%theoretical\cite{philip1972flows,philip1972integral,lauga2003effective,hendy2007effective,teo2009analysis,ng2010apparent}
%and
%experimental\cite{ou2004laminar,ou2005direct,davies2006laminar,choi2006large,maynes2007laminar,lee2008structured,tsai2009quantifying}
%studies have been conducted on those functionalized surfaces and compared to
%the theoretical predictions of the Cassie and Wenzel models\cite{callies2005water}. The
%large interest of the scientific and technological communities on this topic
%is highly motivated by the potential functional applications these surfaces
%have in different realms well-beyond microfluidics, as ship transport, automotive
%and textile industries, electronics, agriculture,
%etc\cite{zhang2008superhydrophobic}.

In literature, the experimental measurements of water slippage past SHS,
usually characterized in terms of effective slip
length\cite{lauga2005microfluidics}, spread over a quite broad data range which
spans from  hundreds of nanometers\cite{joseph2006slippage} to few hundreds of
micrometers\cite{lee2008structured}.  Slight discrepancies between measurements
and theoretical predictions are also reported and mainly ascribed to
experimental factors which cannot always be easily controlled or quantified.  One of those is certainly the
meniscus deformation\cite{richardson1973no,jansons1988determination} resulting from the balance of liquid pressure and surface
tension.
Theoretical\cite{sbragaglia2007note,davis2009geometric,crowdy2010slip} and
numerical\cite{teo2010flow,ng2009stokes,wang2013effects} studies have
demonstrated the negative impact of the meniscus curvature and position
on the effective slip length. Such a detrimental effect has been confirmed by
experimental investigations of SHS characterized via several techniques like
confocal microscopy\cite{ou2004laminar}, micro particle image velocimetry
(\mupiv)\cite{tsai2009quantifying}, pressure versus flow rate measurements\cite{kim2012pressure} 
and dynamic surface force apparatus\cite{steinberger2007high}. 
%It has been demonstrated\cite{richardson1973no,jansons1988determination} that
%meniscus curvature can have detrimental effects in terms of slippage. In the
%experiments reported by Steinberger et al.\cite{steinberger2007high}, a
%mattress of bubbles protruding toward the liquid bulk could deteriorate the
%slippage of a superhydrophobic surface up to the point that the Cassie state
%could show smaller effective slip length than what attained in the Wenzel
%state. 
Similar results were also reported by numerical analysis via continuum
models\cite{teo2009flow}, molecular dynamics\cite{gentili2013water} and Lattice-Boltzmann
simulations\cite{hyvaluoma2011simulations}.  
Very recently, a superhydrophobic microfluidic device with active control of the meniscus shape
has been presented\cite{karatay2013control}. Through numerical and experimental
investigations, the authors showed how the resulting effective slip length
could be tuned and optimized by controlling the protrusion depth of the meniscus into the flow.

%Another phenomenon which might be responsible for dramatic increase in the
%apparent surface friction is the possible contamination of the liquid-air
%interface.  At the typical length scale of microfluidic devices, the high
%surface to volume ratio can result in capillary forces playing a dominant role
%in the flow dynamics.  Particularly, it has been
%reported\cite{ybert1998ascending} that a gradient of concentration of
%contaminant molecules at a water-air interface may generate a shear stress due
%to the Marangoni effect which opposes the water flow nearby the meniscus,
%causing higher friction at the interface.  A similar effect can occur when the
%concentration gradient is replaced with a thermal gradient.  
Another phenomenon which might be responsible for a dramatic increase in the
apparent surface friction, but has not yet received adequate attention in the
study of SHS, is the fact that a water-air interface cannot always be treated
as a stress free boundary.  In that respect, there are quite few experimental
evidences\cite{kim2012pressure,lazouskaya2006interfacial,parkinson2008terminal,manor2008dynamic,manica2009interpreting,zheng2012role,yang2011dynamics}
showing that at the micrometer scale partial-slip or even no-slip assumptions
can be more appropriate, in some circumstances, for modelling water-air
interface under experimental conditions.  Lazouskaya et
al.\cite{lazouskaya2006interfacial} were the first to report a reduced mobility
at the liquid-air interface in an open capillary channel, questioning the
validity of the no-shear boundary condition at the meniscus interface.  More
recently, the same research group performed new experiments with a very similar
geometry\cite{zheng2012role} and highlighted the dependence of an effective
interfacial shear stress on the channel dimensions. Particularly, for channel
width below the water capillary length (i.e. 2.7 mm), the interfacial shear
stress increases as the channel size decreases. Investigating the speed of the
advancing meniscus flooding an open hydrophilic channel, Yang et
al.\cite{yang2011dynamics} found a good agreement between experimental data and
theoretical predictions when a no-slip boundary condition was assumed at the
liquid-air interface. 
%
%COMMENT ON PNAS LOHSE. ADD THERE HAVE BEEN CONFIGURATIONS TO DETECT BOTH
%MENISCUS AND FLOW. THIS ARE LIMITED TO ONE SURFACE PATTERN. WE PROVED TO TRACK
%INTERFACE IN VERY COMPLICATED PATTERNS (ELIPSES). NO WORK PREVIOUSLY USED THIS
%TO PROBE THE LIQUID-AIR INTERFACE BOUNDARY CONDITION.
%

Since the benefits of SHS
in terms of friction reduction rely on the presence of stress-free
boundaries, it is essential to probe whether or not the liquid-air interfaces
can always be considered as a perfect slip interface.
In this paper, we fully characterize the flow past a silicon micro-grooved surface 
and we specifically focus our attention
on the interfacial friction behaviour of the liquid-air interfaces.
However, discriminating between
the effects of a deformed meniscus, on one side, and the
interfacial friction at the liquid-air interface, on the other side, is not an 
easy task as this
requires measuring the flow nearby the meniscus while simultaneously determining
its shape and position. 
Experimental configurations to probe both the liquid flow and the
meniscus deformation has been recently introduced\cite{kim2012pressure,karatay2013control,byun2008direct}. 
For all those studies, the direction normal to the SHS 
has to be co-planar to the microscope focal plane. This necessarily limits the 
experimental investigation to one single surface geometry, that is
the one with micro-grooves perpendicular to the flow direction. Micro-structured surfaces with micro-grooves
parallel to the flow direction or with pillars and holes would be much more difficult to be studied
in that experimental configuration.
In this respect, we recently developed a novel technique\cite{bolognesi2013novel} to
perform simultaneous velocity and interface profile measurements on SHS. That technique
has no restriction on the SHS orientation and it has been successfully used to study even very 
complicated surface pattern like ellipse-shaped pillars.
We now adopt such a technique to characterize the interfacial friction of a
water flow past a silicon surface with longitudinal micro-grooves (see
FIG.~\ref{fig:3dshs}). 
%Since in our experiments we did not implement any control system of the interface
%geometry, the meniscus shape and position spontaneously adapt accordingly to
%the pressure difference between the liquid and gas phases and to the complex
%dynamics of the contact line. As a consequence of that, in our experimental
%conditions the geometry of the liquid-air interfaces may vary both in the
%transverse direction between non-communicating grooves and in the streamwise
%directions due to pressure losses.  
We first investigate the slippage of the SHS globally by measuring 
the effective slip length of the velocity
profile averaged over a surface pitch. We then focus on the local interfacial friction
and we evaluate the friction contributions of the
liquid-solid and liquid-air interfaces, separately.
By correlating the flow with the actual meniscus profile,
we eventually determine what boundary condition best describes the liquid-air interface behaviour.
To best of our knowledge, such a strict test
of the local boundary conditions at the liquid-air interfaces for
micro-structured SHS has never been performed so far. 

%%%%%%  Experimental methods %%%%%%%%%%%%%%%%%%%55
\section{Methods and materials}
\label{sec:methods}
%
%\begin{figure}[t!] \centering
%\includegraphics[angle=-90,width=\columnwidth]{figures/pivsetup.eps}
%\caption{Scheme of the experimental set-up.
%A DPSS (green) laser beam, periodically deflected by an acusto-optic modulator (AOM),
%excites the passive tracers, whose fluorescent light is recorded by the CCD camera sensor.
%} \label{fig:setup}
%\end{figure}
%
In this section, we provide a brief description of the experimental
set-up and procedures, the microfluidic device and the velocity and interface profile
measurement techniques we used for characterizing the slippage at the superhydrophobic
surface. A detailed presentation of those methods are reported elsewhere\cite{bolognesi2013novel}.

\subsection{Experimental set-up}
A Nikon Eclipse TE 2000-U inverted microscope, equipped with a water immersion
$\times60$ (NA $1.2$) objective, is used to capture fluorescence images of the
flow inside the microfluidic chip fed by a flow control system (Fluigent
MFCS-Flex).  A DPSS laser (Cni MLL532, $400$mW at $532\:$nm) excites red
fluorescent $0.3\:\mu$m diameter polystyrene microspheres, dispersed in a
ultrapure Milli-Q water flow ($0.02\%$ solid concentration).  The fluorescent
particles act as passive tracers for tracking both the velocity and the
liquid-air and liquid-solid interface profiles.  A CCD camera (Allied Vision
Technologies) records the light emitted by the tracers at a frame rate of
200Hz. An acousto-optic-modulator (AA Optoelectronics MT80) periodically
deflects the laser beam so that the camera sensor is exposed to the beam light
for $10\:\mu$s only.  

\subsection{Microfluidic chip}
\begin{figure}[t!] \centering
\includegraphics[angle=-90,width=0.7\columnwidth]{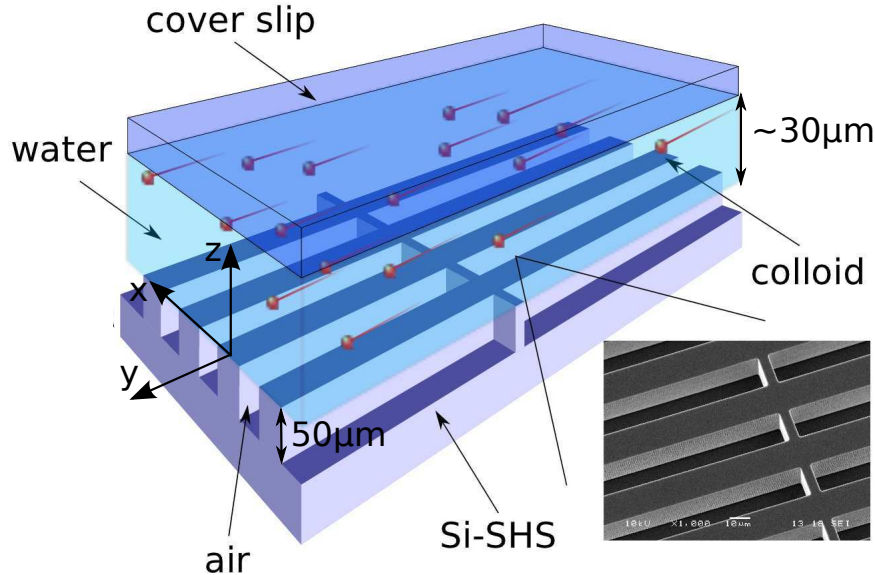}
\caption{Three dimensional schematic view of the microchannel together
with a SEM image of the Si-SHS (inset). 
The ultrapure water and the fluorescent colloids
flow in the direction parallel to the microgrooves, wherein air pockets are
trapped. 
} \label{fig:3dshs} \end{figure}
We used a silicon superhydrophobic surface (Si-SHS), patterned with longitudinal microgrooves,
$16\:\mu$m in width and $50\:\mu$m in depth, evenly spaced by $16\:\mu$m (see FIG.~\ref{fig:3dshs}). 
To prevent in case of a Wenzel transition the
cavities being flooded along the whole channel length, the grooves are sectioned in
smaller grooves $1\:$mm in length, separated each other by transverse $1\:\mu$m thick walls,
as highlighted in the inset of FIG.~\ref{fig:3dshs}.
The Si-SHS is first oxidized by oxygen plasma
treatment and then silanized via standard vapor deposition of
1H,1H,2H,2H-Perfluorooctyltrichlorosilane; 97\% (ABCR) to promote a stable
Cassie state.  The channel is finally assembled by clamping together the Si-SHS 
with a microscope cover slip covered with a PDMS film about $30\:\mu$m
in thickness.  A $2\:$mm$\times40\:$mm area was previously engraved on
the PDMS film.  The resulting channel is then
$40\:$mm long, $2\:$mm wide and $30\:\mu$m deep\cite{bolognesithesis}.
Because of the high aspect-ratio of the cross-section, the confinement effects on the fluid flow
are negligible.

\subsection{Velocity measurements}
Using \mupiv\ cross-correlation analysis of sets of images recorded at
different focal planes\cite{tsai2009quantifying}, we determined
the velocity profile throughout the channel depth. The number $N$ of images per
plane is chosen so that at the minimum detectable speed (about $30\:\mu$m/s) the Brownian motion
component of tracer velocity is negligible (less than $1\%$) with respect to the average 
component. As a consequence of that\cite{bolognesi2013novel}, we recorded $N=4000$
frames at focal planes nearby the silicon surface and $N=3200$ anywhere else.

A reference coordinate system as the one shown in FIG.~\ref{fig:3dshs} is adopted.
The $z$ coordinate axis is parallel to the optical axis
and positively oriented from the surface to the cover slip. Its origin 
is at the liquid-solid interface. The $y$ and $x$ axes are parallel
and transverse to the flow direction, respectively.
The axial step of the objective position is $0.3\:\mu$m in the bulk
and $0.1\:\mu$m close to the plane $z=0$. Processing algorithms are implemented via
custom Matlab codes.
\subsection{Interface detection}
We recently introduced a novel technique\cite{bolognesi2013novel} to measure the
liquid-air and liquid-solid interface relative positions and shapes with a
resolution of tens of nanometers.  Such a measurement is performed by
post-processing the same fluorescent microscopy images used for the \mupiv\ velocity 
analysis. Consequently, the velocity and interface profiles can be
simultaneously determined with a single channel scan.  The method is based on
the detection of the excluded volume, namely the region next to the interface
which is not accessible to the moving colloids. This volume is determined by
splitting the fluorescent images in overlapping rectangular
interrogation windows and, then, measuring for each window the intensity of the
fluorescent light emitted by the flowing tracers as a function of axial
position.  Nearby the interfaces, the intensity versus axial position plot can
be fitted to a shifted error function. The  inflection point of the best-fit
function can be reasonably assumed as the center of those colloids which flowed
closest to the interfaces, namely the boundary of the excluded volume.  By
assembling data from all interrogation windows, a 3D reconstruction of the
excluded volume boundary can be performed in the whole field of view of the
microscope objective. That boundary provides the relative position and the
shape of both liquid-air and liquid-solid interfaces. By considering the
thickness of the depletion layer, the absolute position of both interfaces can
be retrieved with the diameter of the passive tracers being a good estimate 
for the accuracy of such a measurement.

We have also introduced\cite{bolognesi2013novel} an additional
liquid-solid interface detection algorithm, which is in good agreement with
the previous one.  Such a method is capable of measuring the
absolute axial position of a solid wall with nanometer accuracy, provided the
wall reflects the light emitted by the tracers.  This is the case for a silicon
flat surface which, acting as mirror, lets us track both the real and virtual
(reflected) images of the tracers.  The resulting velocity profiles are then
symmetric with respect to the position where flow velocity vanishes to zero.
The solid wall is then detected by locating the axes of
symmetry of those profiles. A fine localization of the solid walls 
is necessary for high quality characterization of the slippage. 
For such a task, we adopt the reflection-based interface detection scheme since
it can locate the absolute position of the liquid-solid interfaces more accurately than the excluded-volume based one.

\subsection{Cleaning and safety procedures}
Standard microfluidic cleaning procedures were adopted for manufacturing the chip
and preparing the solution. Microfluidic chip fabrication and oxygen plasma treatment
were performed in class ISO 7 cleanroom. Silane vapor deposition was done inside
a desiccator placed under a chemical fume hood whereas the chip was assembled on 
a laboratory worktop. The solution of milli-Q water and colloids was prepared and stored
using clean plastic disposable vials, syringe and needles. To prevent the accumulation
of fluorescent colloids on the Si-SHS, the silicon surface was cleaned
via acetone ultrasonic bath and rinsed with DI water before reusing it.
Microfluidic tubings and fittings were frequently cleaned with a mixture of DI water
and ethanol first and then washed with DI water only. For cleaning and safety reasons, gloves were worn at all the time
during sample preparation and experiments.
%%%%%%%%%%%%%%%%%%%%%%%%%%%%%%%%%%%%%%%%%%%%%%%%%%%%%%%%%%%%%%%%%%%
%%%%%%%%%%%%%% Results and Discussion %%%%%%%%%%%%%%%%%%%55%%%%%%%
%%%%%%%%%%%%%%%%%%%%%%%%%%%%%%%%%%%%%%%%%%%%%%%%%%%%%%%%%%%%%%%%%%%

\section{Results and discussion}
%Definizione di effective slip length
Slippage occurs at liquid-gas and liquid-solid interfaces whenever the
tangential component of the liquid velocity at the interface appears to be
different from the one of the other phase (either solid or gas). That velocity
gap between the two phases is called slip velocity.  A well known quantity
which has been used to characterize the slippage is the slip length $b$ defined
as the ratio between the slip velocity and the shear rate at the
interface\cite{lauga2005microfluidics}.  When the slip length is experimentally
estimated by averaging an appropriate measurement over a length scale much
larger than the molecular scale, we rather refer to it as effective slip
length.  For structured surfaces consisting of a periodic array of liquid-air
and liquid-solid interfaces, if the measurement is averaged over one array
pitch, we refer to it as global effective slip length. Conversely, if the slip
length is measured just on a pitch fraction, like the liquid-air or the
liquid-solid interface only, we denote it as local effective slip length.  In
the next sections, we characterize the slippage of the examined Si-SHS by
determining both the global and the local effective slip lengths.
\subsection{Global effective slip length} \label{subsec:global}
\begin{figure}[t!] \centering
\includegraphics[angle=-90,width=\columnwidth]{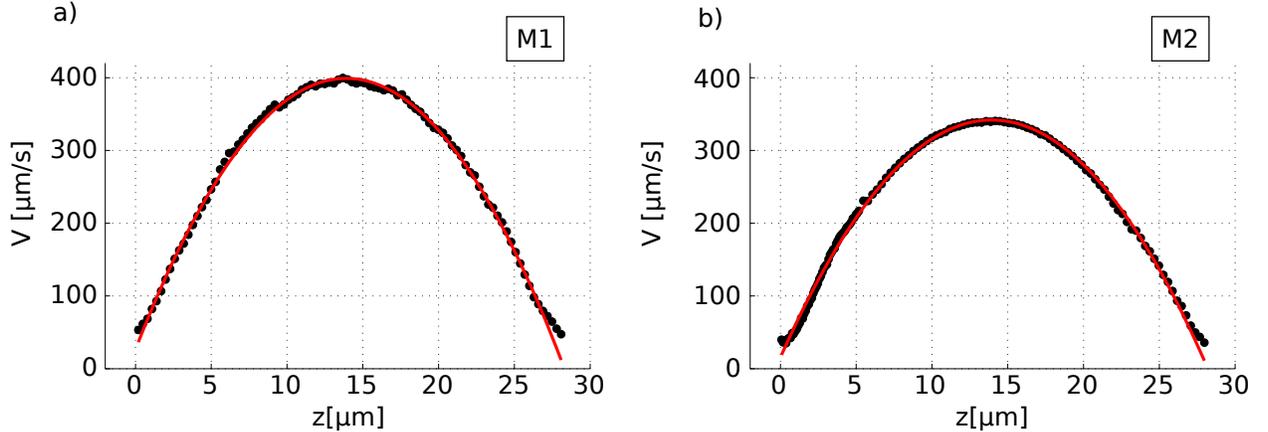}
\caption{Streamwise velocity profiles (solid circles) averaged over the
roughness pitch for measurement M1 (left panel) and measurement M2 (right panel), together with the best-fit
functions (solid lines).  } \label{fig:vel_prof} \end{figure}

We first assess the overall performance of our device by measuring the global
effective slip length at the Si-SHS.  Fluorescent images are acquired with a
scan along the channel depth from $z=-6\:\mu$m to $z=30\:\mu$m.  For sake of
simplicity, we recall that the origin of the $z$ axis is fixed at the
liquid-solid interface of the Si-SHS.  The streamwise velocity profile is
averaged over the microscope field of view in the flow (y) direction and over a
surface pitch, consisting of one liquid-solid and one liquid-air area, in the
transverse (x) direction.  The resulting profile is fitted to that one
$\overline{V}(z)$ predicted by the one-dimensional (1D) model of a pressure
driven flow with a sticky wall at $z=H$ (i.e. the cover slip) and a partially
slippery wall at $z=0$ (i.e. the Si-SHS) with global effective slip length
$b_{gl}$. The parameter $H$ stands for the channel depth.  The analytical
expression for $\overline{V}(z)$ reads
\begin{equation} \overline{V}(z) = 4 V_{p}  \frac{ (b_{gl} + H) (H - z)  \left[
b_{gl} (z + H) + z H  \right] } {  H^2 (2b_{gl} + H)^2  }
\label{eq:partial_slip} \end{equation} 
where $V_p$ is the peak velocity. The fit parameters are $b_{gl}$, $H$ and
$V_{p}$. It is worth noting that the channel depth parameter $H$ has to be
fitted because we could not apply the reflection-based liquid-solid detection
technique at the cover slip wall z=$H$.  Indeed the reflection coefficient of
glass is too low to accurately track the virtual images of the colloids flowing
next to the glass wall. Additionally, due to the elasticity of PDMS, the
channel depth might change between experiments as it depends on the clamping
force which keeps together the PDMS-coated cover slip and the Si-SHS.  The
measurements were performed in two different spots of the device. We refers to
those measurements with the labels M1 and M2.  Between M1 and M2, the
microchannel was disassembled, cleaned and reassembled.  FIG.~\ref{fig:vel_prof}
shows the experimental profiles $\overline{V}(z)$ and the corresponding
best-fit curves.  The best-fit parameters are reported in TABLE \ref{tab:param}
together with the fit uncertainties. It is worth recalling that in our system
we do not control the geometry of the liquid-air interface and thus the
meniscus can freely adapt its shape and position according to the local
experimental conditions. As a consequence of that, there is no reason for the
menisci to share the same shapes and positions between measurements M1 and M2.
\begin{table}[t!] \centering \begin{tabular}{|c|c|c|} \hline & \bf M1 &  \bf M2
\\ \hline $V_{p}$ & $399.3\pm0.5\:\mu$m/s  & $341.4\pm0.5\:\mu$m/s  \\ $H$ &
$28.31\pm0.04\:\mu$m  & $28.24\pm0.04\:\mu$m   \\ $b_{gl}$ &
$0.49\pm\:0.05\:\mu$m & $0.33\pm0.04\:\mu$m \\ \hline \end{tabular}
\caption{Best-fit parameters and corresponding uncertainties for the 1D model,
see Eq.(\ref{eq:partial_slip}), for measurements M1 and M2: peak velocity
$V_p$, channel depth $H$, global effective slip length $b_{gl}$.  }
\label{tab:param} \end{table}
%
%The precision of the measured values is given by the uncertainty of the
%corresponding fitting parameters.  For better estimating the precision of the
%channel depth and global effective slip length measurements, we added to the
%fit uncertainties the resolution of the reflection-based solid wall detection.

To better assess the effectiveness of the friction reduction at the
Si-SHS, we compare the experimental results to the theoretical model of a surface
with a periodic pattern of no-slip and no-shear stripes parallel to the flow\cite{philip1972flows}.
The predicted value for the global effective slip length is 
\begin{equation}
b_{th} = - \frac{L}{\pi} \log \left[ \cos  \left( \frac{\pi}{2}
(1 - \Phi_s) \right)  \right ]
\label{eq:philip}
\end{equation}
where $L$ is the pattern pitch and $\Phi_s$ the solid fraction, the latter being defined as
the ratio between the no-slip stripe width and the pattern pitch.
 %we have $\Phi_s=0.5$ and $L=32\:\mu$m. According to Eq.(\ref{eq:philip}),
%Considering a Si-SHS with a solid fraction of 0.5 and a surface pitch of $32\:\mu$m,
For the examined Si-SHS whose solid fraction is $\Phi_s=0.5$ and surface pitch $L=32\:\mu$m,
the expected global effective slip length is $b_{th}=3.53\: \mu$m, namely ten times
larger than what we actually measured. Moreover, we remark that the Si-SHS sample shows highly
inhomogeneous slippage as the global effective slip length varies up to $40\%$ between M1 and M2.

\subsection{Local effective slip length}
In order to clarify this apparent contradiction between experimental and theoretical
results, a local investigation of the surface friction at the liquid-air interfaces is required.
Indeed the meniscus curvature at the surface cavity depends on the pressure difference
between the air and liquid phases as well as the surface tension according to
the Laplace law. These elements together with the meniscus position, dictated
by the complex dynamics of the contact line, can be significantly affected by
often uncontrolled parameters (e.g.  surface defects, impurities) and thus
highly varies along different regions of the Si-SHS samples. That could in principle
justify the measured inhomogeneity of surface slippage as well as the strong
increase of the surface friction. Nonetheless, a reduced mobility on a flat
liquid-air interface could also produce the same effects. In order to discriminate
between those two scenarios, we rely on our excluded-volume based
interface detection technique to measure the local interfacial friction
at both the liquid-solid and liquid-air interfaces. 
We first determine the interface profiles
and then we correlate them to the corresponding flows. 
%Indeed, by taking into account the actual
%shape of that liquid-air interface and analysing the flow past that curved
%meniscus, we can measure the interfacial friction between liquid and air phases
%and thus determine what is the boundary condition which accurately describes
%the behaviour of the interface.  

%
\begin{figure}[t!] \centering
\includegraphics[angle=-90,width=0.8\columnwidth]{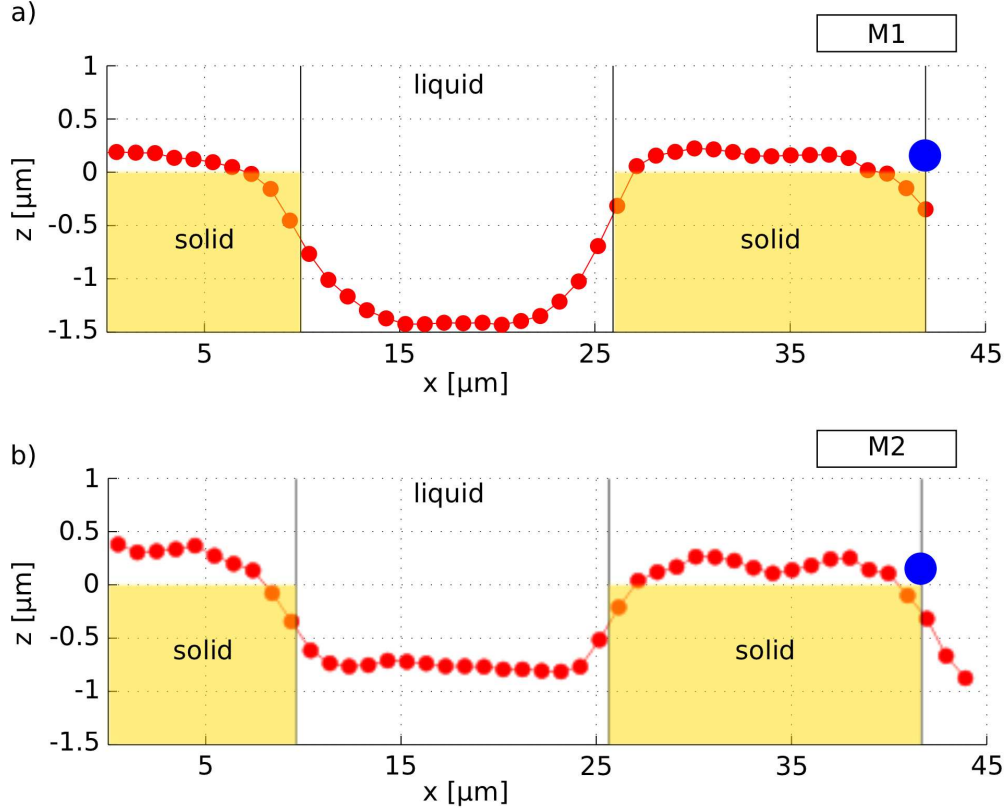}
\caption{
Excluded volume boundary profiles along the transverse direction
$x$ for M1 (top panel) and M2 (bottom panel). The passive tracer
size along the axial direction $z$ is represented by the largest solid circles.} 
\label{fig:int_prof} \end{figure}
FIG.~\ref{fig:int_prof} shows the transverse profiles of the boundaries of
the excluded volumes for both M1 and M2.  As reported
elsewhere\cite{bolognesi2013novel}, the absolute position of the measured
boundaries are accurate within a tracer diameter (namely, $0.3\:\mu$m), whose
size is also shown in figure for comparison.  The distance between the actual
interfaces and the excluded volume boundary depends on the thickness of the
tracer depletion layer.  At the solid surface, the depletion layer is manly due
to the impenetrability of the moving colloids with either the surface or
possible tracers stuck to it. Thus the depletion layer thickness may vary
approximately between 0.5 to 1.5 particle diameter (namely, between
$0.15\:\mu$m to $0.45\:\mu$m). Other phenomena which might affect the depletion
layer, as the hydrodynamic Saffman lift or electrostatic interactions, can be
neglected in our experiments\cite{bolognesi2013novel}. At the liquid-air
interface, the depletion layer thickness depends on the tracer wettability.
Since polystyrene is an hydrophobic material, particles tend to stuck at the
liquid-air interface\cite{paunov2003novel} with a contact angle of about
$90^\circ$. As a consequence of that, the depletion layer vanishes and the
measured excluded-volume boundary well approximate the meniscus profile.
FIG.~\ref{fig:int_prof} shows how both menisci are almost flat and
penetrates with different depths inside the cavities.  When the meniscus is
modelled as a no-shear interface, partial flooding of the surface cavities
reduces the surface effective slippage and higher filling levels corresponds to smaller
effective slip length\cite{steinberger2007high,gentili2013water} .  Our
experiments show opposite behaviour with higher filling level (that is M1)
corresponding to higher slippage. That is already an indirect evidence that
perfect-slip condition may not apply to the examined liquid-air interfaces.

\begin{figure}[t!] \centering
\includegraphics[angle=-90,width=0.6\columnwidth]{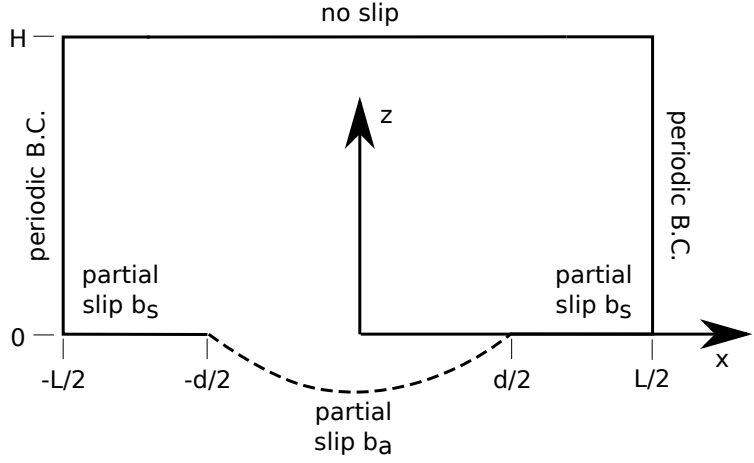}
\caption{2D domain and boundary conditions where the best-fit function
$V_{th}(x,y)$ is solved for. The dashed line represents the meniscus transverse
profile measured through the excluded volume based method (cf.
FIG.~\ref{fig:int_prof}).  The parameter $d=L/2$ stands for the liquid-air and
liquid-solid interface width.  } \label{fig:computationalvolume} \end{figure}

\begin{figure}[t!] \centering
\includegraphics[angle=-90,width=\columnwidth]{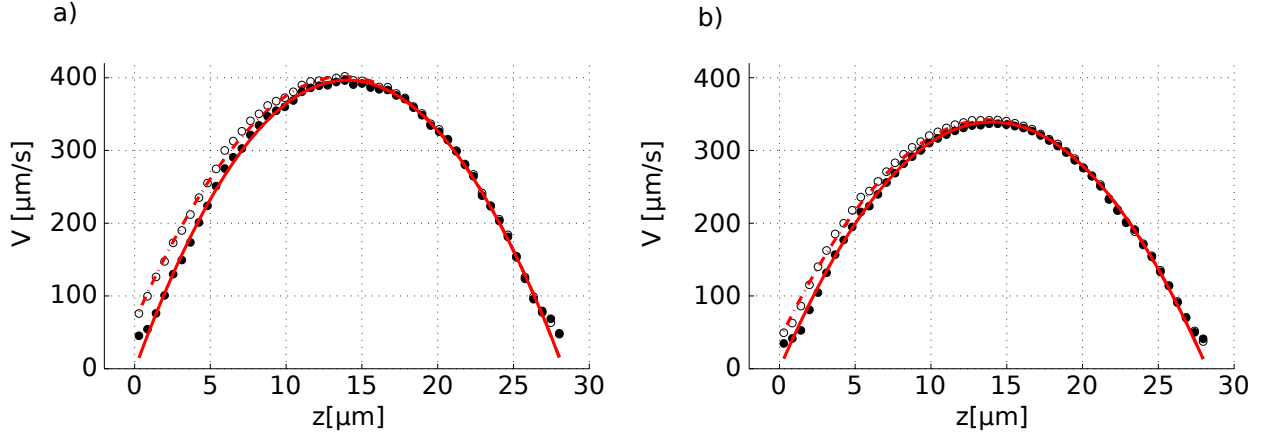}
\caption{Streamwise velocity profiles at the middle point of the liquid-air
(empty circles) and liquid-solid (solid circles) interfaces for M1 (left panel)
and M2 (right panel). The dashed and the solid lines are the corresponding
best-fit profiles.  } \label{fig:loc_vel_prof} \end{figure}

To probe what is the actual boundary condition needed to effectively model the
flow past the measured liquid-air interface, we first average the streamwise
velocity field along the flow direction $y$. The resulting experimental
discrete function $V(x,z)$ is fitted to the two-dimensional (2D) model
$V_{th}(x,z)$ which is solution of the Poisson equation:
\begin{equation} \nabla^2 V_{th}(x,z)  = \nabla p / \mu \\ \label{eq:Poisson2D}
\end{equation}
with the following boundary conditions
\begin{equation} \begin{cases} V_{th}(-L/2,z)=V_{th}(L/2,z) \\ \frac{\partial
V_{th}}{\partial x}|_{x=\pm L/2} = 0 \\ V_{th}(xy) = b_s \frac{\partial
V_{th}}{ \partial z} \quad \text{for} \quad  (x,y)\in \text{liquid-solid
interface}  \\ V_{th}(x,y) = b_a \frac{\partial V_{th}}{ \partial z} \quad
\text{for} \quad (x,y)\in \text{liquid-air interface} \\ V_{th}(x,H) = 0 \\
\end{cases} \label{eq:BC} \end{equation}
where $p$ is the liquid pressure and $\mu$ the fluid dynamic viscosity.  The
domain where Eq.(\ref{eq:Poisson2D}) is solved for is shown in
FIG.~\ref{fig:computationalvolume} together with the boundary conditions. In
such a model, the meniscus is shaped according to the liquid-air interface
profile measurements reported in FIG.~\ref{fig:int_prof}, whereas the
liquid-solid interfaces, located  at $z=0$, are assumed to be flat.
Since we cannot find an analytical closed-form for $V_{th}(x,z)$, we used
Comsol Multiphysics$^{\text{TM}}$ to perform the fit between the experimental
2D profile $V(x,z)$ and the numerical solution of the 2D theoretical model
$V_{th}(x,z)$.  The fit parameters are the ratio $\alpha=\nabla p /\mu$ between
the pressure gradient and the dynamic viscosity, the channel depth $H$ and the
local effective slip lengths $b_s$ and $b_a$.  The latter parameter is a direct
measurement of the interfacial friction at the actual liquid-air interface.
The experimental and best-fit velocity profiles are shown in
FIG.~\ref{fig:loc_vel_prof}. The solid and empty circles represent the
experimental profiles at the middle point of the liquid-solid and liquid-air
interfaces, respectively. Similarly, the solid and the dashed lines are given
by the best-fit function $V_{th}(x,z)$ evaluated at the middle points of the
liquid-solid and liquid-air interfaces, respectively. The best-fit parameters
are shown in TABLE~\ref{tab:param3}.  A good estimate for the accuracy of the
local effective slip length measurements is given by the passive tracer
diameter.  We can use the best-fit function $V_{th}(x,z)$ to calculate the
interfacial shear stresses at the middle point of both the liquid-solid and
liquid-air interfaces as follows \begin{equation} \tau = \mu \nabla V_{th}
\cdot \mathbf{n} \label{eq:tau} \end{equation} where $\mathbf{n}$ is the unit
vector normal to the interface. The results are shown in TABLE~\ref{tab:param4}, where
$\tau_s$ and $\tau_a$ refer to the liquid-solid and liquid-air interfaces,
respectively.
\begin{table}[t!] \centering \begin{tabular}{|c|c|c|} \hline & \bf M1 &  \bf M2
\\ \hline $\alpha$ & $3.86\pm0.01\:(\mu$m$\cdot$s$)^{-1}$  & $3.33\pm0.01\:$
$(\mu$m$\cdot$s$)^{-1}$ \\ $H$ & $28.31\pm0.05\:\mu$m  &  $28.24\pm0.05\:\mu$m
\\ $b_{s}$ & $-0.02\pm0.15\:\mu$m & $0.00\pm0.15\:\mu$m  \\ $b_{a}$ &
$-0.10\pm0.15\:\mu$m & $0.09\pm0.15\:\mu$m  \\ \hline \end{tabular}
\caption{Best-fit parameters for
the 2D model, see Eq.(\ref{eq:Poisson2D}), for M1 and M2: pressure gradient and
dynamic viscosity ratio $\alpha$, channel depth $H$, local effective slip
lengths $b_{s}$ and $b_{a}$.  } \label{tab:param3} \end{table}
\begin{table}[t!] \centering \begin{tabular}{|c|c|c|} \hline & \bf M1 &  \bf M2
\\ \hline $\tau_s$ & $60\pm2\:$mPa & $48\pm2\:$mPa\\  \hline $\tau_a$ &
$50\pm5\:$mPa & $45\pm5\:$mPa\\  \hline \end{tabular} \caption{ Interfacial
shear stresses at the middle point of the liquid-solid and liquid-air
interfaces for M1 and M2, calculated according to Eq.(\ref{eq:tau}).  }
\label{tab:param4} \end{table}
%
%To probe what is actual boundary condition needed to effectively model the
%flow past the actual liquid-air interface, we fit again the experimental
%velocity V(x,y) to Eq.(\ref{eq:Poisson2D}) using Comsol
%Multiphysics$^{\text{TM}}$.  We now consider a slightly different model
%domain. The nominal flat meniscus is now replaced with the measured liquid-air
%interface profile reported in Fig.\ref{fig:int_prof}, whereas the liquid-solid
%interfaces, located  at $z=0$, stay flat as shown in
%Fig.\ref{fig:computationalvolume}b.  The boundary conditions are the same as
%Eq.(\ref{eq:BC}) but a partial slippage $\hat{b}_a$ is now assumed at the
%meniscus protruding within the cavity.  The fit parameter $\hat{b}_a$ is here
%referred to as the adjusted local effective slip length.  That parameter is a
%direct measurement of the partial slippage at the actual liquid-air interface.
%Since the effective slip length was always measured with respect to an
%arbitrary reference level (usually, the liquid-solid interface), it was not
%possible to separate the contributions on the surface slippage due to
%different effects as a partial slippage at the liquid-air interface on one
%side and the meniscus deformation on the other.
%
%Values of the adjusted local slip length for the liquid-air interface and the
%local slip length for liquid-solid interfaces are reported in
%Tab.\ref{tab:param3}. 
For both M1 and M2, non-zero interfacial shear stresses appear
at the menisci, which turn out to be, within the accuracy of the measurements,
closer to no-slip interfaces rather than no-shear ones.  
%We can thus
%conclude that for the examined Si-SHS, shear-free boundary conditions do not
%apply at the liquid-air interface. 

\subsection{Discussion}
Performing simultaneous interface and velocity
profile measurements with our technique
reveals that the assumption of a no-shear boundary condition at the liquid-air interface
is not valid for
the examined SHS. That is the reason why the predicted
global effective slippage of  Eq.(\ref{eq:philip})
is much higher than the measured values. 
If instead we consider a model where the stress-free boundary
condition is replaced with a low partial-slip (namely, $b_a\ll L$),
the corresponding global effective slip length
can be estimated through the following heuristic formula\cite{ybert2007achieving}  
\begin{equation}
b_{th} = (1-\Phi_s) b_a
\label{eq:ybert}
\end{equation}
According to this model, for $\Phi_s=0.5$ the expected global slip length is $b_a/2$
whereas we measured values of $b_{gl}$ larger than $b_a$
(cf. TABLE~\ref{tab:param} and TABLE~\ref{tab:param3}). 
In order to explain the observed behaviour,
we must take into account the actual profile of the menisci. FIG.~\ref{fig:int_prof} shows
that the menisci are essentially flat. Consequently, in our experiments
any detrimental effect induced
by the meniscus curvature\cite{steinberger2007high,karatay2013control} on the global slippage is negligible.
On the other hand, the menisci are not co-planar with the
liquid-solid interfaces as it is assumed in the model,
but they partially penetrate the surface cavities.
Since the liquid-air interfaces are almost immobilised (i.e. $b_a\simeq 0$), the
partial flooding of the cavities is responsible for an apparent
positive global effective slippage, namely $b_{gl}>0$.
Had been the meniscus directed towards the flow, we would have 
measured a negative global effective slip length.
Additionally, the fact that the position of the menisci can vary along the surface sample
accounts for the observation of an apparent inhomogeneity of the surface slippage.
The difference between global effective slippages in M1 and M2 is about 40\%
and so it is the difference between the corresponding penetration depths of the menisci.

The possible causes behind the onset of an interfacial friction
at the liquid-air interface for the examined sample are several. The presence of contaminant particles within the flow
is one of those. More specifically, the
liquid-air interface is confined in a $1\:$mm$\times16\:\mu$m rectangular
region, consisting of two adjacent grooves and two consecutive transverse
$1\mu$m thick walls (see FIG.~\ref{fig:3dshs}).   
Consequently, contaminant particles could be adsorbed onto the liquid-air interface
and pushed by the liquid flow against the downstream transverse wall. That would result in a 
particle concentration gradient and, possibly, Marangoni stresses
which oppose the water flow nearby the meniscus,
causing higher friction at the interface\cite{ybert1998ascending}.
Possible agents acting as surface contaminants could be either 
PDMS molecules coming from the channel walls or unknown
surfactant molecules used by the supplier in the preparation
of the fluorescent polystyrene colloids or even the colloids themselves.
In that respect, it is worth noting that
experiments involving the use of SHS are usually
performed following standard microfluidic cleaning procedures.
As detailed in the section~\ref{sec:methods}, we followed those procedures  and we did not 
implement any other specialized cleaning protocol 
to control the level of contamination of the liquid-air interfaces.
Alternative to the surface contamination scenario, 
we remind that Marangoni stresses could also raise from \mupiv\ laser-induced thermal
effects. A detailed investigation about the actual causes behind the measured
interfacial friction at the meniscus are beyond the scope of this research.

\section{Conclusions}
In the present paper, we applied our novel velocity and interface detection
technique to thoroughly characterize the slippage behaviour of a silicon
micro-grooved superhydrophobic surface. We first investigated the global
slippage capability by averaging the velocity profiles along a periodic array
consisting of one liquid-air and one liquid-solid interface. 
%The position of
%the silicon walls, with respect to which the global slippage was measured, were
%determined with few nanometers accuracy via a reflection-based detection
%method, we recently introduced in a previous work.  
The surface performances were much lower than
expected when modelling the surface as a periodic pattern of co-planar no-shear
and no-slip parallel stripes. In addition, the surface shows an inhomogeneous
behaviour when slippage is probed in different spots of the samples. To better
clarify the discrepancy with the theory, we evaluated the interfacial friction at
the liquid-solid and liquid-air interfaces by measuring the meniscus profiles and
relating them to the flow nearby. 
In doing so, we were able to probe the actual boundary condition
at the deformed meniscus and we find out that the liquid-air interface is 
not behaving like a no-shear boundary, as it is usually assumed in the studies
of SHS. Unexpectedly, the global slippage
of the examined Si-SHS is a mere consequence of almost no-slip menisci protruding
within the surface cavities. Due to the impossibility to control the meniscus geometry, 
the global slippage varies along the sample  according to the local shape and position of the menisci.
%Inspired by similar cases reported in literature of apparent friction at the liquid-air interfaces, we 
%considered possible explanations to justify a non-vanishing stress at the
%meniscus interfaces. Further investigations will follow to corroborate or reject
%the proposed scenarios.

In conclusion, we reported for the first time the violation under specific
circumstances of the widely common assumption of stress-free boundary at the
liquid-air interfaces in micro-structured SHS. We showed that even in presence
of a stable Cassie state, the friction
reduction capabilities of SHS can be seriously compromised when interfacial
stresses appears at the liquid-air interfaces.  We proved that those potential
detrimental effects in terms of low slippage can be as worse as those
deriving from a deformed meniscus protruding into the
flow\cite{steinberger2007high,karatay2013control}. 
Consequently, the presence of non-deformed (i.e. flat) liquid-air interfaces
in the cavities of a SHS can no longer be considered a sufficient condition to guarantee significant
surface friction reduction.

\section{Acknowledgements}
We are pleased to thank L. Bocquet for interesting discussion
and  C. Ybert for both interesting discussions and help with the development
of the experimental set-up. We also thank the
Universit\`a Italo Francese and the French embassy in Italy for their financial support
and the Lyon Institute of Nanotechnology (INL) for the use of their technological
facilities. This work was partly supported by the French RENATECH network.

% If you have acknowledgments, this puts in the proper section head.
%\begin{acknowledgments}
% Put your acknowledgments here.
%\end{acknowledgments}

% Create the reference section using BibTeX:
%\bibliography{manuscript}

\begin{thebibliography}{38}%
\makeatletter
\providecommand \@ifxundefined [1]{%
 \@ifx{#1\undefined}
}%
\providecommand \@ifnum [1]{%
 \ifnum #1\expandafter \@firstoftwo
 \else \expandafter \@secondoftwo
 \fi
}%
\providecommand \@ifx [1]{%
 \ifx #1\expandafter \@firstoftwo
 \else \expandafter \@secondoftwo
 \fi
}%
\providecommand \natexlab [1]{#1}%
\providecommand \enquote  [1]{``#1''}%
\providecommand \bibnamefont  [1]{#1}%
\providecommand \bibfnamefont [1]{#1}%
\providecommand \citenamefont [1]{#1}%
\providecommand \href@noop [0]{\@secondoftwo}%
\providecommand \href [0]{\begingroup \@sanitize@url \@href}%
\providecommand \@href[1]{\@@startlink{#1}\@@href}%
\providecommand \@@href[1]{\endgroup#1\@@endlink}%
\providecommand \@sanitize@url [0]{\catcode `\\12\catcode `\$12\catcode
  `\&12\catcode `\#12\catcode `\^12\catcode `\_12\catcode `\%12\relax}%
\providecommand \@@startlink[1]{}%
\providecommand \@@endlink[0]{}%
\providecommand \url  [0]{\begingroup\@sanitize@url \@url }%
\providecommand \@url [1]{\endgroup\@href {#1}{\urlprefix }}%
\providecommand \urlprefix  [0]{URL }%
\providecommand \Eprint [0]{\href }%
\providecommand \doibase [0]{http://dx.doi.org/}%
\providecommand \selectlanguage [0]{\@gobble}%
\providecommand \bibinfo  [0]{\@secondoftwo}%
\providecommand \bibfield  [0]{\@secondoftwo}%
\providecommand \translation [1]{[#1]}%
\providecommand \BibitemOpen [0]{}%
\providecommand \bibitemStop [0]{}%
\providecommand \bibitemNoStop [0]{.\EOS\space}%
\providecommand \EOS [0]{\spacefactor3000\relax}%
\providecommand \BibitemShut  [1]{\csname bibitem#1\endcsname}%
\let\auto@bib@innerbib\@empty
%</preamble>
\bibitem [{\citenamefont {Rothstein}(2010)}]{rothstein2010slip}%
  \BibitemOpen
  \bibfield  {author} {\bibinfo {author} {\bibfnamefont {J.}~\bibnamefont
  {Rothstein}},\ }\bibfield  {title} {\enquote {\bibinfo {title} {Slip on
  superhydrophobic surfaces},}\ }\href@noop {} {\bibfield  {journal} {\bibinfo
  {journal} {Annual Review of Fluid Mechanics}\ }\textbf {\bibinfo {volume}
  {42}},\ \bibinfo {pages} {89--109} (\bibinfo {year} {2010})}\BibitemShut
  {NoStop}%
\bibitem [{\citenamefont {Cottin-Bizonne}\ \emph {et~al.}(2003)\citenamefont
  {Cottin-Bizonne}, \citenamefont {Barrat}, \citenamefont {Bocquet},\ and\
  \citenamefont {Charlaix}}]{cottin2003low}%
  \BibitemOpen
  \bibfield  {author} {\bibinfo {author} {\bibfnamefont {C.}~\bibnamefont
  {Cottin-Bizonne}}, \bibinfo {author} {\bibfnamefont {J.-L.}\ \bibnamefont
  {Barrat}}, \bibinfo {author} {\bibfnamefont {L.}~\bibnamefont {Bocquet}}, \
  and\ \bibinfo {author} {\bibfnamefont {E.}~\bibnamefont {Charlaix}},\
  }\bibfield  {title} {\enquote {\bibinfo {title} {Low-friction flows of liquid
  at nanopatterned interfaces},}\ }\href@noop {} {\bibfield  {journal}
  {\bibinfo  {journal} {Nature materials}\ }\textbf {\bibinfo {volume} {2}},\
  \bibinfo {pages} {237--240} (\bibinfo {year} {2003})}\BibitemShut {NoStop}%
\bibitem [{\citenamefont {Peters}\ \emph {et~al.}(2009)\citenamefont {Peters},
  \citenamefont {Pirat}, \citenamefont {Sbragaglia}, \citenamefont {Borkent},
  \citenamefont {Wessling}, \citenamefont {Lohse},\ and\ \citenamefont
  {Lammertink}}]{peters2009cassie}%
  \BibitemOpen
  \bibfield  {author} {\bibinfo {author} {\bibfnamefont {A.}~\bibnamefont
  {Peters}}, \bibinfo {author} {\bibfnamefont {C.}~\bibnamefont {Pirat}},
  \bibinfo {author} {\bibfnamefont {M.}~\bibnamefont {Sbragaglia}}, \bibinfo
  {author} {\bibfnamefont {B.}~\bibnamefont {Borkent}}, \bibinfo {author}
  {\bibfnamefont {M.}~\bibnamefont {Wessling}}, \bibinfo {author}
  {\bibfnamefont {D.}~\bibnamefont {Lohse}}, \ and\ \bibinfo {author}
  {\bibfnamefont {R.}~\bibnamefont {Lammertink}},\ }\bibfield  {title}
  {\enquote {\bibinfo {title} {Cassie-baxter to wenzel state wetting
  transition: Scaling of the front velocity},}\ }\href@noop {} {\bibfield
  {journal} {\bibinfo  {journal} {The European Physical Journal E}\ }\textbf
  {\bibinfo {volume} {29}},\ \bibinfo {pages} {391--397} (\bibinfo {year}
  {2009})}\BibitemShut {NoStop}%
\bibitem [{\citenamefont {Giacomello}\ \emph
  {et~al.}(2012{\natexlab{a}})\citenamefont {Giacomello}, \citenamefont
  {Meloni}, \citenamefont {Chinappi},\ and\ \citenamefont
  {Casciola}}]{giacomello2012cassie}%
  \BibitemOpen
  \bibfield  {author} {\bibinfo {author} {\bibfnamefont {A.}~\bibnamefont
  {Giacomello}}, \bibinfo {author} {\bibfnamefont {S.}~\bibnamefont {Meloni}},
  \bibinfo {author} {\bibfnamefont {M.}~\bibnamefont {Chinappi}}, \ and\
  \bibinfo {author} {\bibfnamefont {C.~M.}\ \bibnamefont {Casciola}},\
  }\bibfield  {title} {\enquote {\bibinfo {title} {Cassie--baxter and wenzel
  states on a nanostructured surface: Phase diagram, metastabilities, and
  transition mechanism by atomistic free energy calculations},}\ }\href@noop {}
  {\bibfield  {journal} {\bibinfo  {journal} {Langmuir}\ }\textbf {\bibinfo
  {volume} {28}},\ \bibinfo {pages} {10764--10772} (\bibinfo {year}
  {2012}{\natexlab{a}})}\BibitemShut {NoStop}%
\bibitem [{\citenamefont {Giacomello}\ \emph
  {et~al.}(2012{\natexlab{b}})\citenamefont {Giacomello}, \citenamefont
  {Chinappi}, \citenamefont {Meloni},\ and\ \citenamefont
  {Casciola}}]{giacomello2012metastable}%
  \BibitemOpen
  \bibfield  {author} {\bibinfo {author} {\bibfnamefont {A.}~\bibnamefont
  {Giacomello}}, \bibinfo {author} {\bibfnamefont {M.}~\bibnamefont
  {Chinappi}}, \bibinfo {author} {\bibfnamefont {S.}~\bibnamefont {Meloni}}, \
  and\ \bibinfo {author} {\bibfnamefont {C.~M.}\ \bibnamefont {Casciola}},\
  }\bibfield  {title} {\enquote {\bibinfo {title} {Metastable wetting on
  superhydrophobic surfaces: Continuum and atomistic views of the
  cassie-baxter--wenzel transition},}\ }\href@noop {} {\bibfield  {journal}
  {\bibinfo  {journal} {Physical review letters}\ }\textbf {\bibinfo {volume}
  {109}},\ \bibinfo {pages} {226102} (\bibinfo {year}
  {2012}{\natexlab{b}})}\BibitemShut {NoStop}%
\bibitem [{\citenamefont {Lauga}, \citenamefont {Brenner},\ and\ \citenamefont
  {Stone}(2005)}]{lauga2005microfluidics}%
  \BibitemOpen
  \bibfield  {author} {\bibinfo {author} {\bibfnamefont {E.}~\bibnamefont
  {Lauga}}, \bibinfo {author} {\bibfnamefont {M.}~\bibnamefont {Brenner}}, \
  and\ \bibinfo {author} {\bibfnamefont {H.}~\bibnamefont {Stone}},\ }\bibfield
   {title} {\enquote {\bibinfo {title} {Microfluidics: The no-slip boundary
  condition},}\ }\href@noop {} {\bibfield  {journal} {\bibinfo  {journal}
  {Fluid Dynamics}\ ,\ \bibinfo {pages} {1--27}} (\bibinfo {year}
  {2005})}\BibitemShut {NoStop}%
\bibitem [{\citenamefont {Joseph}\ \emph {et~al.}(2006)\citenamefont {Joseph},
  \citenamefont {Cottin-Bizonne}, \citenamefont {Benoit}, \citenamefont
  {Ybert}, \citenamefont {Journet}, \citenamefont {Tabeling},\ and\
  \citenamefont {Bocquet}}]{joseph2006slippage}%
  \BibitemOpen
  \bibfield  {author} {\bibinfo {author} {\bibfnamefont {P.}~\bibnamefont
  {Joseph}}, \bibinfo {author} {\bibfnamefont {C.}~\bibnamefont
  {Cottin-Bizonne}}, \bibinfo {author} {\bibfnamefont {J.}~\bibnamefont
  {Benoit}}, \bibinfo {author} {\bibfnamefont {C.}~\bibnamefont {Ybert}},
  \bibinfo {author} {\bibfnamefont {C.}~\bibnamefont {Journet}}, \bibinfo
  {author} {\bibfnamefont {P.}~\bibnamefont {Tabeling}}, \ and\ \bibinfo
  {author} {\bibfnamefont {L.}~\bibnamefont {Bocquet}},\ }\bibfield  {title}
  {\enquote {\bibinfo {title} {Slippage of water past superhydrophobic carbon
  nanotube forests in microchannels},}\ }\href@noop {} {\bibfield  {journal}
  {\bibinfo  {journal} {Physical review letters}\ }\textbf {\bibinfo {volume}
  {97}},\ \bibinfo {pages} {156104} (\bibinfo {year} {2006})}\BibitemShut
  {NoStop}%
\bibitem [{\citenamefont {Lee}, \citenamefont {Choi},\ and\ \citenamefont
  {Kim}(2008)}]{lee2008structured}%
  \BibitemOpen
  \bibfield  {author} {\bibinfo {author} {\bibfnamefont {C.}~\bibnamefont
  {Lee}}, \bibinfo {author} {\bibfnamefont {C.}~\bibnamefont {Choi}}, \ and\
  \bibinfo {author} {\bibfnamefont {C.}~\bibnamefont {Kim}},\ }\bibfield
  {title} {\enquote {\bibinfo {title} {Structured surfaces for a giant liquid
  slip},}\ }\href@noop {} {\bibfield  {journal} {\bibinfo  {journal} {Physical
  review letters}\ }\textbf {\bibinfo {volume} {101}},\ \bibinfo {pages}
  {64501} (\bibinfo {year} {2008})}\BibitemShut {NoStop}%
\bibitem [{\citenamefont {Richardson}(1973)}]{richardson1973no}%
  \BibitemOpen
  \bibfield  {author} {\bibinfo {author} {\bibfnamefont {S.}~\bibnamefont
  {Richardson}},\ }\bibfield  {title} {\enquote {\bibinfo {title} {On the
  no-slip boundary condition},}\ }\href@noop {} {\bibfield  {journal} {\bibinfo
   {journal} {Journal of Fluid Mechanics}\ }\textbf {\bibinfo {volume} {59}},\
  \bibinfo {pages} {707--719} (\bibinfo {year} {1973})}\BibitemShut {NoStop}%
\bibitem [{\citenamefont {Jansons}(1988)}]{jansons1988determination}%
  \BibitemOpen
  \bibfield  {author} {\bibinfo {author} {\bibfnamefont {K.}~\bibnamefont
  {Jansons}},\ }\bibfield  {title} {\enquote {\bibinfo {title} {Determination
  of the macroscopic (partial) slip boundary condition for a viscous flow over
  a randomly rough surface with a perfect slip microscopic boundary
  condition},}\ }\href@noop {} {\bibfield  {journal} {\bibinfo  {journal}
  {Physics of Fluids}\ }\textbf {\bibinfo {volume} {31}},\ \bibinfo {pages}
  {15} (\bibinfo {year} {1988})}\BibitemShut {NoStop}%
\bibitem [{\citenamefont {Sbragaglia}\ and\ \citenamefont
  {Prosperetti}(2007)}]{sbragaglia2007note}%
  \BibitemOpen
  \bibfield  {author} {\bibinfo {author} {\bibfnamefont {M.}~\bibnamefont
  {Sbragaglia}}\ and\ \bibinfo {author} {\bibfnamefont {A.}~\bibnamefont
  {Prosperetti}},\ }\bibfield  {title} {\enquote {\bibinfo {title} {A note on
  the effective slip properties for microchannel flows with ultrahydrophobic
  surfaces},}\ }\href@noop {} {\bibfield  {journal} {\bibinfo  {journal}
  {Physics of Fluids}\ }\textbf {\bibinfo {volume} {19}},\ \bibinfo {pages}
  {043603} (\bibinfo {year} {2007})}\BibitemShut {NoStop}%
\bibitem [{\citenamefont {Davis}\ and\ \citenamefont
  {Lauga}(2009)}]{davis2009geometric}%
  \BibitemOpen
  \bibfield  {author} {\bibinfo {author} {\bibfnamefont {A.}~\bibnamefont
  {Davis}}\ and\ \bibinfo {author} {\bibfnamefont {E.}~\bibnamefont {Lauga}},\
  }\bibfield  {title} {\enquote {\bibinfo {title} {Geometric transition in
  friction for flow over a bubble mattress},}\ }\href@noop {} {\bibfield
  {journal} {\bibinfo  {journal} {Physics of Fluids}\ }\textbf {\bibinfo
  {volume} {21}},\ \bibinfo {pages} {011701} (\bibinfo {year}
  {2009})}\BibitemShut {NoStop}%
\bibitem [{\citenamefont {Crowdy}(2010)}]{crowdy2010slip}%
  \BibitemOpen
  \bibfield  {author} {\bibinfo {author} {\bibfnamefont {D.}~\bibnamefont
  {Crowdy}},\ }\bibfield  {title} {\enquote {\bibinfo {title} {Slip length for
  longitudinal shear flow over a dilute periodic mattress of protruding
  bubbles},}\ }\href@noop {} {\bibfield  {journal} {\bibinfo  {journal}
  {Physics of Fluids}\ }\textbf {\bibinfo {volume} {22}},\ \bibinfo {pages}
  {121703} (\bibinfo {year} {2010})}\BibitemShut {NoStop}%
\bibitem [{\citenamefont {Teo}\ and\ \citenamefont
  {Khoo}(2010{\natexlab{a}})}]{teo2010flow}%
  \BibitemOpen
  \bibfield  {author} {\bibinfo {author} {\bibfnamefont {C.}~\bibnamefont
  {Teo}}\ and\ \bibinfo {author} {\bibfnamefont {B.}~\bibnamefont {Khoo}},\
  }\bibfield  {title} {\enquote {\bibinfo {title} {Flow past superhydrophobic
  surfaces containing longitudinal grooves: effects of interface curvature},}\
  }\href@noop {} {\bibfield  {journal} {\bibinfo  {journal} {Microfluidics and
  Nanofluidics}\ }\textbf {\bibinfo {volume} {9}},\ \bibinfo {pages} {499--511}
  (\bibinfo {year} {2010}{\natexlab{a}})}\BibitemShut {NoStop}%
\bibitem [{\citenamefont {Ng}\ and\ \citenamefont {Wang}(2009)}]{ng2009stokes}%
  \BibitemOpen
  \bibfield  {author} {\bibinfo {author} {\bibfnamefont {C.}~\bibnamefont
  {Ng}}\ and\ \bibinfo {author} {\bibfnamefont {C.}~\bibnamefont {Wang}},\
  }\bibfield  {title} {\enquote {\bibinfo {title} {Stokes shear flow over a
  grating: Implications for superhydrophobic slip},}\ }\href@noop {} {\bibfield
   {journal} {\bibinfo  {journal} {Physics of Fluids}\ }\textbf {\bibinfo
  {volume} {21}},\ \bibinfo {pages} {013602} (\bibinfo {year}
  {2009})}\BibitemShut {NoStop}%
\bibitem [{\citenamefont {Wang}, \citenamefont {Teo},\ and\ \citenamefont
  {Khoo}()}]{wang2013effects}%
  \BibitemOpen
  \bibfield  {author} {\bibinfo {author} {\bibfnamefont {L.}~\bibnamefont
  {Wang}}, \bibinfo {author} {\bibfnamefont {C.}~\bibnamefont {Teo}}, \ and\
  \bibinfo {author} {\bibfnamefont {B.}~\bibnamefont {Khoo}},\ }\bibfield
  {title} {\enquote {\bibinfo {title} {Effects of interface deformation on flow
  through microtubes containing superhydrophobic surfaces with longitudinal
  ribs and grooves},}\ }\href@noop {} {\bibinfo  {journal} {Microfluidics and
  Nanofluidics}\ ,\ \bibinfo {pages} {DOI
  10.1007/s10404--013--1201--1}}\BibitemShut {NoStop}%
\bibitem [{\citenamefont {Ou}, \citenamefont {Perot},\ and\ \citenamefont
  {Rothstein}(2004)}]{ou2004laminar}%
  \BibitemOpen
\bibfield  {journal} {  }\bibfield  {author} {\bibinfo {author} {\bibfnamefont
  {J.}~\bibnamefont {Ou}}, \bibinfo {author} {\bibfnamefont {B.}~\bibnamefont
  {Perot}}, \ and\ \bibinfo {author} {\bibfnamefont {J.}~\bibnamefont
  {Rothstein}},\ }\bibfield  {title} {\enquote {\bibinfo {title} {Laminar drag
  reduction in microchannels using ultrahydrophobic surfaces},}\ }\href@noop {}
  {\bibfield  {journal} {\bibinfo  {journal} {Physics of fluids}\ }\textbf
  {\bibinfo {volume} {16}},\ \bibinfo {pages} {4635} (\bibinfo {year}
  {2004})}\BibitemShut {NoStop}%
\bibitem [{\citenamefont {Tsai}\ \emph {et~al.}(2009)\citenamefont {Tsai},
  \citenamefont {Peters}, \citenamefont {Pirat}, \citenamefont {Wessling},
  \citenamefont {Lammertink},\ and\ \citenamefont
  {Lohse}}]{tsai2009quantifying}%
  \BibitemOpen
  \bibfield  {author} {\bibinfo {author} {\bibfnamefont {P.}~\bibnamefont
  {Tsai}}, \bibinfo {author} {\bibfnamefont {A.}~\bibnamefont {Peters}},
  \bibinfo {author} {\bibfnamefont {C.}~\bibnamefont {Pirat}}, \bibinfo
  {author} {\bibfnamefont {M.}~\bibnamefont {Wessling}}, \bibinfo {author}
  {\bibfnamefont {R.}~\bibnamefont {Lammertink}}, \ and\ \bibinfo {author}
  {\bibfnamefont {D.}~\bibnamefont {Lohse}},\ }\bibfield  {title} {\enquote
  {\bibinfo {title} {Quantifying effective slip length over micropatterned
  hydrophobic surfaces},}\ }\href@noop {} {\bibfield  {journal} {\bibinfo
  {journal} {Physics of Fluids}\ }\textbf {\bibinfo {volume} {21}},\ \bibinfo
  {pages} {112002} (\bibinfo {year} {2009})}\BibitemShut {NoStop}%
\bibitem [{\citenamefont {Kim}\ and\ \citenamefont
  {Hidrovo}(2012)}]{kim2012pressure}%
  \BibitemOpen
  \bibfield  {author} {\bibinfo {author} {\bibfnamefont {T.}~\bibnamefont
  {Kim}}\ and\ \bibinfo {author} {\bibfnamefont {C.}~\bibnamefont {Hidrovo}},\
  }\bibfield  {title} {\enquote {\bibinfo {title} {Pressure and partial wetting
  effects on superhydrophobic friction reduction in microchannel flow},}\
  }\href@noop {} {\bibfield  {journal} {\bibinfo  {journal} {Physics of
  Fluids}\ }\textbf {\bibinfo {volume} {24}},\ \bibinfo {pages}
  {112003--112003} (\bibinfo {year} {2012})}\BibitemShut {NoStop}%
\bibitem [{\citenamefont {Steinberger}\ \emph {et~al.}(2007)\citenamefont
  {Steinberger}, \citenamefont {Cottin-Bizonne}, \citenamefont {Kleimann},\
  and\ \citenamefont {Charlaix}}]{steinberger2007high}%
  \BibitemOpen
  \bibfield  {author} {\bibinfo {author} {\bibfnamefont {A.}~\bibnamefont
  {Steinberger}}, \bibinfo {author} {\bibfnamefont {C.}~\bibnamefont
  {Cottin-Bizonne}}, \bibinfo {author} {\bibfnamefont {P.}~\bibnamefont
  {Kleimann}}, \ and\ \bibinfo {author} {\bibfnamefont {E.}~\bibnamefont
  {Charlaix}},\ }\bibfield  {title} {\enquote {\bibinfo {title} {High friction
  on a bubble mattress},}\ }\href@noop {} {\bibfield  {journal} {\bibinfo
  {journal} {Nature Materials}\ }\textbf {\bibinfo {volume} {6}},\ \bibinfo
  {pages} {665--668} (\bibinfo {year} {2007})}\BibitemShut {NoStop}%
\bibitem [{\citenamefont {Teo}\ and\ \citenamefont
  {Khoo}(2010{\natexlab{b}})}]{teo2009flow}%
  \BibitemOpen
  \bibfield  {author} {\bibinfo {author} {\bibfnamefont {C.}~\bibnamefont
  {Teo}}\ and\ \bibinfo {author} {\bibfnamefont {B.}~\bibnamefont {Khoo}},\
  }\bibfield  {title} {\enquote {\bibinfo {title} {Flow past superhydrophobic
  surfaces containing longitudinal grooves: effects of interface curvature},}\
  }\href@noop {} {\bibfield  {journal} {\bibinfo  {journal} {Microfluidics and
  Nanofluidics}\ }\textbf {\bibinfo {volume} {9}},\ \bibinfo {pages} {499--511}
  (\bibinfo {year} {2010}{\natexlab{b}})}\BibitemShut {NoStop}%
\bibitem [{\citenamefont {Gentili}\ \emph {et~al.}(2013)\citenamefont
  {Gentili}, \citenamefont {Chinappi}, \citenamefont {Bolognesi}, \citenamefont
  {Giacomello},\ and\ \citenamefont {Casciola}}]{gentili2013water}%
  \BibitemOpen
  \bibfield  {author} {\bibinfo {author} {\bibfnamefont {D.}~\bibnamefont
  {Gentili}}, \bibinfo {author} {\bibfnamefont {M.}~\bibnamefont {Chinappi}},
  \bibinfo {author} {\bibfnamefont {G.}~\bibnamefont {Bolognesi}}, \bibinfo
  {author} {\bibfnamefont {A.}~\bibnamefont {Giacomello}}, \ and\ \bibinfo
  {author} {\bibfnamefont {C.}~\bibnamefont {Casciola}},\ }\bibfield  {title}
  {\enquote {\bibinfo {title} {Water slippage on hydrophobic nanostructured
  surfaces: molecular dynamics results for different filling levels},}\
  }\href@noop {} {\bibfield  {journal} {\bibinfo  {journal} {Meccanica}\ ,\
  \bibinfo {pages} {1--9}} (\bibinfo {year} {2013})}\BibitemShut {NoStop}%
\bibitem [{\citenamefont {Hyv{\"a}luoma}, \citenamefont {Kunert},\ and\
  \citenamefont {Harting}(2011)}]{hyvaluoma2011simulations}%
  \BibitemOpen
  \bibfield  {author} {\bibinfo {author} {\bibfnamefont {J.}~\bibnamefont
  {Hyv{\"a}luoma}}, \bibinfo {author} {\bibfnamefont {C.}~\bibnamefont
  {Kunert}}, \ and\ \bibinfo {author} {\bibfnamefont {J.}~\bibnamefont
  {Harting}},\ }\bibfield  {title} {\enquote {\bibinfo {title} {Simulations of
  slip flow on nanobubble-laden surfaces},}\ }\href@noop {} {\bibfield
  {journal} {\bibinfo  {journal} {Journal of Physics: Condensed Matter}\
  }\textbf {\bibinfo {volume} {23}},\ \bibinfo {pages} {184106} (\bibinfo
  {year} {2011})}\BibitemShut {NoStop}%
\bibitem [{\citenamefont {Karatay}\ \emph {et~al.}(2013)\citenamefont
  {Karatay}, \citenamefont {Haase}, \citenamefont {Visser}, \citenamefont
  {Sun}, \citenamefont {Lohse}, \citenamefont {Tsai},\ and\ \citenamefont
  {Lammertink}}]{karatay2013control}%
  \BibitemOpen
  \bibfield  {author} {\bibinfo {author} {\bibfnamefont {E.}~\bibnamefont
  {Karatay}}, \bibinfo {author} {\bibfnamefont {A.~S.}\ \bibnamefont {Haase}},
  \bibinfo {author} {\bibfnamefont {C.~W.}\ \bibnamefont {Visser}}, \bibinfo
  {author} {\bibfnamefont {C.}~\bibnamefont {Sun}}, \bibinfo {author}
  {\bibfnamefont {D.}~\bibnamefont {Lohse}}, \bibinfo {author} {\bibfnamefont
  {P.~A.}\ \bibnamefont {Tsai}}, \ and\ \bibinfo {author} {\bibfnamefont
  {R.~G.}\ \bibnamefont {Lammertink}},\ }\bibfield  {title} {\enquote {\bibinfo
  {title} {Control of slippage with tunable bubble mattresses},}\ }\href@noop
  {} {\bibfield  {journal} {\bibinfo  {journal} {Proceedings of the National
  Academy of Sciences}\ }\textbf {\bibinfo {volume} {110}},\ \bibinfo {pages}
  {8422--8426} (\bibinfo {year} {2013})}\BibitemShut {NoStop}%
\bibitem [{\citenamefont {Lazouskaya}, \citenamefont {Jin},\ and\ \citenamefont
  {Or}(2006)}]{lazouskaya2006interfacial}%
  \BibitemOpen
  \bibfield  {author} {\bibinfo {author} {\bibfnamefont {V.}~\bibnamefont
  {Lazouskaya}}, \bibinfo {author} {\bibfnamefont {Y.}~\bibnamefont {Jin}}, \
  and\ \bibinfo {author} {\bibfnamefont {D.}~\bibnamefont {Or}},\ }\bibfield
  {title} {\enquote {\bibinfo {title} {Interfacial interactions and colloid
  retention under steady flows in a capillary channel},}\ }\href@noop {}
  {\bibfield  {journal} {\bibinfo  {journal} {Journal of colloid and interface
  science}\ }\textbf {\bibinfo {volume} {303}},\ \bibinfo {pages} {171--184}
  (\bibinfo {year} {2006})}\BibitemShut {NoStop}%
\bibitem [{\citenamefont {Parkinson}\ \emph {et~al.}(2008)\citenamefont
  {Parkinson}, \citenamefont {Sedev}, \citenamefont {Fornasiero},\ and\
  \citenamefont {Ralston}}]{parkinson2008terminal}%
  \BibitemOpen
  \bibfield  {author} {\bibinfo {author} {\bibfnamefont {L.}~\bibnamefont
  {Parkinson}}, \bibinfo {author} {\bibfnamefont {R.}~\bibnamefont {Sedev}},
  \bibinfo {author} {\bibfnamefont {D.}~\bibnamefont {Fornasiero}}, \ and\
  \bibinfo {author} {\bibfnamefont {J.}~\bibnamefont {Ralston}},\ }\bibfield
  {title} {\enquote {\bibinfo {title} {The terminal rise velocity of 10--100
  $\mu$m diameter bubbles in water},}\ }\href@noop {} {\bibfield  {journal}
  {\bibinfo  {journal} {Journal of colloid and interface science}\ }\textbf
  {\bibinfo {volume} {322}},\ \bibinfo {pages} {168--172} (\bibinfo {year}
  {2008})}\BibitemShut {NoStop}%
\bibitem [{\citenamefont {Manor}\ \emph {et~al.}(2008)\citenamefont {Manor},
  \citenamefont {Vakarelski}, \citenamefont {Stevens}, \citenamefont {Grieser},
  \citenamefont {Dagastine},\ and\ \citenamefont {Chan}}]{manor2008dynamic}%
  \BibitemOpen
  \bibfield  {author} {\bibinfo {author} {\bibfnamefont {O.}~\bibnamefont
  {Manor}}, \bibinfo {author} {\bibfnamefont {I.~U.}\ \bibnamefont
  {Vakarelski}}, \bibinfo {author} {\bibfnamefont {G.~W.}\ \bibnamefont
  {Stevens}}, \bibinfo {author} {\bibfnamefont {F.}~\bibnamefont {Grieser}},
  \bibinfo {author} {\bibfnamefont {R.~R.}\ \bibnamefont {Dagastine}}, \ and\
  \bibinfo {author} {\bibfnamefont {D.~Y.}\ \bibnamefont {Chan}},\ }\bibfield
  {title} {\enquote {\bibinfo {title} {Dynamic forces between bubbles and
  surfaces and hydrodynamic boundary conditions},}\ }\href@noop {} {\bibfield
  {journal} {\bibinfo  {journal} {Langmuir}\ }\textbf {\bibinfo {volume}
  {24}},\ \bibinfo {pages} {11533--11543} (\bibinfo {year} {2008})}\BibitemShut
  {NoStop}%
\bibitem [{\citenamefont {Manica}\ \emph {et~al.}(2009)\citenamefont {Manica},
  \citenamefont {Parkinson}, \citenamefont {Ralston},\ and\ \citenamefont
  {Chan}}]{manica2009interpreting}%
  \BibitemOpen
  \bibfield  {author} {\bibinfo {author} {\bibfnamefont {R.}~\bibnamefont
  {Manica}}, \bibinfo {author} {\bibfnamefont {L.}~\bibnamefont {Parkinson}},
  \bibinfo {author} {\bibfnamefont {J.}~\bibnamefont {Ralston}}, \ and\
  \bibinfo {author} {\bibfnamefont {D.~Y.}\ \bibnamefont {Chan}},\ }\bibfield
  {title} {\enquote {\bibinfo {title} {Interpreting the dynamic interaction
  between a very small rising bubble and a hydrophilic titania surface},}\
  }\href@noop {} {\bibfield  {journal} {\bibinfo  {journal} {The Journal of
  Physical Chemistry C}\ }\textbf {\bibinfo {volume} {114}},\ \bibinfo {pages}
  {1942--1946} (\bibinfo {year} {2009})}\BibitemShut {NoStop}%
\bibitem [{\citenamefont {Zheng}\ \emph {et~al.}(2012)\citenamefont {Zheng},
  \citenamefont {Wang}, \citenamefont {Or}, \citenamefont {Lazouskaya},\ and\
  \citenamefont {Jin}}]{zheng2012role}%
  \BibitemOpen
  \bibfield  {author} {\bibinfo {author} {\bibfnamefont {W.}~\bibnamefont
  {Zheng}}, \bibinfo {author} {\bibfnamefont {L.}~\bibnamefont {Wang}},
  \bibinfo {author} {\bibfnamefont {D.}~\bibnamefont {Or}}, \bibinfo {author}
  {\bibfnamefont {V.}~\bibnamefont {Lazouskaya}}, \ and\ \bibinfo {author}
  {\bibfnamefont {Y.}~\bibnamefont {Jin}},\ }\bibfield  {title} {\enquote
  {\bibinfo {title} {The role of mixed boundaries on flow in open capillary
  channels with curved air-water interfaces},}\ }\href@noop {} {\bibfield
  {journal} {\bibinfo  {journal} {Langmuir}\ } (\bibinfo {year}
  {2012})}\BibitemShut {NoStop}%
\bibitem [{\citenamefont {Yang}\ \emph {et~al.}(2011)\citenamefont {Yang},
  \citenamefont {Krasowska}, \citenamefont {Priest}, \citenamefont {Popescu},\
  and\ \citenamefont {Ralston}}]{yang2011dynamics}%
  \BibitemOpen
  \bibfield  {author} {\bibinfo {author} {\bibfnamefont {D.}~\bibnamefont
  {Yang}}, \bibinfo {author} {\bibfnamefont {M.}~\bibnamefont {Krasowska}},
  \bibinfo {author} {\bibfnamefont {C.}~\bibnamefont {Priest}}, \bibinfo
  {author} {\bibfnamefont {M.~N.}\ \bibnamefont {Popescu}}, \ and\ \bibinfo
  {author} {\bibfnamefont {J.}~\bibnamefont {Ralston}},\ }\bibfield  {title}
  {\enquote {\bibinfo {title} {Dynamics of capillary-driven flow in open
  microchannels},}\ }\href@noop {} {\bibfield  {journal} {\bibinfo  {journal}
  {The Journal of Physical Chemistry C}\ }\textbf {\bibinfo {volume} {115}},\
  \bibinfo {pages} {18761--18769} (\bibinfo {year} {2011})}\BibitemShut
  {NoStop}%
\bibitem [{\citenamefont {Byun}\ \emph {et~al.}(2008)\citenamefont {Byun},
  \citenamefont {Kim}, \citenamefont {Ko},\ and\ \citenamefont
  {Park}}]{byun2008direct}%
  \BibitemOpen
  \bibfield  {author} {\bibinfo {author} {\bibfnamefont {D.}~\bibnamefont
  {Byun}}, \bibinfo {author} {\bibfnamefont {J.}~\bibnamefont {Kim}}, \bibinfo
  {author} {\bibfnamefont {H.}~\bibnamefont {Ko}}, \ and\ \bibinfo {author}
  {\bibfnamefont {H.}~\bibnamefont {Park}},\ }\bibfield  {title} {\enquote
  {\bibinfo {title} {Direct measurement of slip flows in superhydrophobic
  microchannels with transverse grooves},}\ }\href@noop {} {\bibfield
  {journal} {\bibinfo  {journal} {Physics of Fluids}\ }\textbf {\bibinfo
  {volume} {20}},\ \bibinfo {pages} {113601} (\bibinfo {year}
  {2008})}\BibitemShut {NoStop}%
\bibitem [{\citenamefont {Bolognesi}\ \emph {et~al.}(2013)\citenamefont
  {Bolognesi}, \citenamefont {Cottin-Bizonne}, \citenamefont {Guene},
  \citenamefont {Teisseire},\ and\ \citenamefont {Pirat}}]{bolognesi2013novel}%
  \BibitemOpen
  \bibfield  {author} {\bibinfo {author} {\bibfnamefont {G.}~\bibnamefont
  {Bolognesi}}, \bibinfo {author} {\bibfnamefont {C.}~\bibnamefont
  {Cottin-Bizonne}}, \bibinfo {author} {\bibfnamefont {E.}~\bibnamefont
  {Guene}}, \bibinfo {author} {\bibfnamefont {J.}~\bibnamefont {Teisseire}}, \
  and\ \bibinfo {author} {\bibfnamefont {C.}~\bibnamefont {Pirat}},\ }\bibfield
   {title} {\enquote {\bibinfo {title} {A novel technique for simultaneous
  velocity and interface profile measurements on micro-structured surfaces},}\
  }\href@noop {} {\bibfield  {journal} {\bibinfo  {journal} {Soft Matter}\
  }\textbf {\bibinfo {volume} {9}},\ \bibinfo {pages} {2239--2244} (\bibinfo
  {year} {2013})}\BibitemShut {NoStop}%
\bibitem [{\citenamefont {Bolognesi}(2012)}]{bolognesithesis}%
  \BibitemOpen
  \bibfield  {author} {\bibinfo {author} {\bibfnamefont {G.}~\bibnamefont
  {Bolognesi}},\ }\href@noop {} {\emph {\bibinfo {title} {Optical studies of
  micron-scale flows}}}\ (\bibinfo  {publisher} {Lap Lambert Academic
  Publishing},\ \bibinfo {year} {2012})\BibitemShut {NoStop}%
\bibitem [{\citenamefont {Philip}(1972)}]{philip1972flows}%
  \BibitemOpen
  \bibfield  {author} {\bibinfo {author} {\bibfnamefont {J.}~\bibnamefont
  {Philip}},\ }\bibfield  {title} {\enquote {\bibinfo {title} {Flows satisfying
  mixed no-slip and no-shear conditions},}\ }\href@noop {} {\bibfield
  {journal} {\bibinfo  {journal} {Zeitschrift f{\"u}r Angewandte Mathematik und
  Physik (ZAMP)}\ }\textbf {\bibinfo {volume} {23}},\ \bibinfo {pages}
  {353--372} (\bibinfo {year} {1972})}\BibitemShut {NoStop}%
\bibitem [{\citenamefont {Paunov}(2003)}]{paunov2003novel}%
  \BibitemOpen
  \bibfield  {author} {\bibinfo {author} {\bibfnamefont {V.~N.}\ \bibnamefont
  {Paunov}},\ }\bibfield  {title} {\enquote {\bibinfo {title} {Novel method for
  determining the three-phase contact angle of colloid particles adsorbed at
  air-water and oil-water interfaces},}\ }\href@noop {} {\bibfield  {journal}
  {\bibinfo  {journal} {Langmuir}\ }\textbf {\bibinfo {volume} {19}},\ \bibinfo
  {pages} {7970--7976} (\bibinfo {year} {2003})}\BibitemShut {NoStop}%
\bibitem [{\citenamefont {Ybert}\ and\ \citenamefont
  {Di~Meglio}(1998)}]{ybert1998ascending}%
  \BibitemOpen
  \bibfield  {author} {\bibinfo {author} {\bibfnamefont {C.}~\bibnamefont
  {Ybert}}\ and\ \bibinfo {author} {\bibfnamefont {J.}~\bibnamefont
  {Di~Meglio}},\ }\bibfield  {title} {\enquote {\bibinfo {title} {Ascending air
  bubbles in protein solutions},}\ }\href@noop {} {\bibfield  {journal}
  {\bibinfo  {journal} {The European Physical Journal B-Condensed Matter and
  Complex Systems}\ }\textbf {\bibinfo {volume} {4}},\ \bibinfo {pages}
  {313--319} (\bibinfo {year} {1998})}\BibitemShut {NoStop}%
\bibitem [{\citenamefont {Ybert}\ \emph {et~al.}(2007)\citenamefont {Ybert},
  \citenamefont {Barentin}, \citenamefont {Cottin-Bizonne}, \citenamefont
  {Joseph},\ and\ \citenamefont {Bocquet}}]{ybert2007achieving}%
  \BibitemOpen
  \bibfield  {author} {\bibinfo {author} {\bibfnamefont {C.}~\bibnamefont
  {Ybert}}, \bibinfo {author} {\bibfnamefont {C.}~\bibnamefont {Barentin}},
  \bibinfo {author} {\bibfnamefont {C.}~\bibnamefont {Cottin-Bizonne}},
  \bibinfo {author} {\bibfnamefont {P.}~\bibnamefont {Joseph}}, \ and\ \bibinfo
  {author} {\bibfnamefont {L.}~\bibnamefont {Bocquet}},\ }\bibfield  {title}
  {\enquote {\bibinfo {title} {Achieving large slip with superhydrophobic
  surfaces: Scaling laws for generic geometries},}\ }\href@noop {} {\bibfield
  {journal} {\bibinfo  {journal} {Physics of fluids}\ }\textbf {\bibinfo
  {volume} {19}},\ \bibinfo {pages} {123601} (\bibinfo {year}
  {2007})}\BibitemShut {NoStop}%
\bibitem [{\citenamefont {Chaudhury}\ and\ \citenamefont
  {Whitesides}(1991)}]{chaudhury1991direct}%
  \BibitemOpen
  \bibfield  {author} {\bibinfo {author} {\bibfnamefont {M.~K.}\ \bibnamefont
  {Chaudhury}}\ and\ \bibinfo {author} {\bibfnamefont {G.~M.}\ \bibnamefont
  {Whitesides}},\ }\bibfield  {title} {\enquote {\bibinfo {title} {Direct
  measurement of interfacial interactions between semispherical lenses and flat
  sheets of poly (dimethylsiloxane) and their chemical derivatives},}\
  }\href@noop {} {\bibfield  {journal} {\bibinfo  {journal} {Langmuir}\
  }\textbf {\bibinfo {volume} {7}},\ \bibinfo {pages} {1013--1025} (\bibinfo
  {year} {1991})}\BibitemShut {NoStop}%
\end{thebibliography}

\providecommand{\noopsort}[1]{}\providecommand{\singleletter}[1]{#1}%

\end{document}